\documentclass[sigconf]{acmart}

\AtBeginDocument{%
  \providecommand\BibTeX{{%
    \normalfont B\kern-0.5em{\scshape i\kern-0.25em b}\kern-0.8em\TeX}}}

\setcopyright{rightsretained}
\copyrightyear{2021}
\acmYear{2021}
\acmDOI{}
\acmISBN{} 

\begin{document}

\title[]{Location Data Reveals Disproportionate Disaster Impact Amongst the Poor: A Case Study of the 2017 Puebla Earthquake Using \textit{Mobilkit}}

\author{Takahiro Yabe}
\email{tyabe@purdue.edu}
\affiliation{%
  \institution{Purdue University, World Bank}
  \country{USA}
}
\author{Nicholas K. W. Jones}
\email{njones@worldbank.org}
\affiliation{%
  \institution{World Bank}
    \country{USA}
}
\author{Nancy Lozano-Gracia}
\email{nlozano@worldbank.org}
\affiliation{%
  \institution{World Bank}
    \country{USA}
}
\author{Maham Faisal Khan}
\email{mkhan57@worldbank.org}
\affiliation{%
  \institution{World Bank}
    \country{USA}
}
\author{Satish V. Ukkusuri}
\email{sukkusur@purdue.edu}
\affiliation{%
  \institution{Purdue University, World Bank}
    \country{USA}
}
\author{Samuel Fraiberger}
\email{sfraiberger@worldbank.org}
\affiliation{%
  \institution{World Bank}
    \country{USA}
}
\author{Aleister Montfort}
\email{amontfortibieta@worldbank.org}
\affiliation{%
  \institution{World Bank}
    \country{Mexico}
}

\renewcommand{\shortauthors}{Yabe et al.}

\begin{abstract}
Location data obtained from smartphones is increasingly finding use cases in disaster risk management. Where traditionally, CDR has provided the predominant digital footprint for human mobility, GPS data now has immense potential in terms of improved spatiotemporal accuracy, volume, availability, and accessibility. GPS data has already proven invaluable in a range of pre- and post-disaster use cases, such as quantifying displacement, measuring rates of return and recovery, evaluating accessibility to critical resources, planning for resilience. Despite its popularity and potential, however, the use of GPS location data in DRM is still nascent, with several use cases yet to be explored. In this paper, we consider the 2017 Puebla Earthquake in Mexico to (i) validate and expand upon post-disaster analysis applications using GPS data, and (ii) illustrate the use of a new toolkit, \textit{Mobilkit}, to facilitate scalable, replicable extensions of this work for a wide range of disasters, including earthquakes, typhoons, flooding, and beyond.
\end{abstract}



\maketitle

\section{Introduction}

As global temperatures increase, intense climate-related disasters rise, and the world grapples with economic losses at an unprecedented scale owing to the coronavirus pandemic, it has never been more important to foster sustainable development and resilience in areas prone to hazards \cite{hallegatte2016unbreakable}. Traditionally, when these risks have materialized, their impact has been quantified using census data and surveys---measures that are costly and challenging to scale  \cite{fussell2014recovery}. More recently, however, location datasets made available by technology companies and cellular providers have made it possible to access and measure high resolution changes to human mobility quickly and at relatively low cost.

Initially, these primarily took the form of Call Detail Records (CDRs), which triangulate user locations from network towers. CDR data features in several well-proven disaster use cases \cite{bengtsson2015using,finger2016mobile,bengtsson2011improved,wilson2016rapid}. However, access depends on case-by-case negotiations with providers, and is only spatially precise within the `cell' where a user is triangulated. As smartphones equipped with GPS technologies become more ubiquitous, location data traces have become far more resolved than CDR data, enabling mapping of user trajectories in space (precise latitude and longitude) and time (down to seconds) \cite{gonzalez2008understanding,blondel2015survey}. GPS traces have the potential to power a new wave of resilience and disaster risk management (DRM) efforts. Notably, Yabe et al. \cite{yaberecovery} recently leveraged GPS data to understand population displacement and return patterns in the contexts of five disasters across Puerto Rico, Florida, and Japan. GPS data also typically required negotiation with fewer providers with penetration in several countries at a time.

\begin{figure*}[t]
\centering
\includegraphics[width=.8\linewidth]{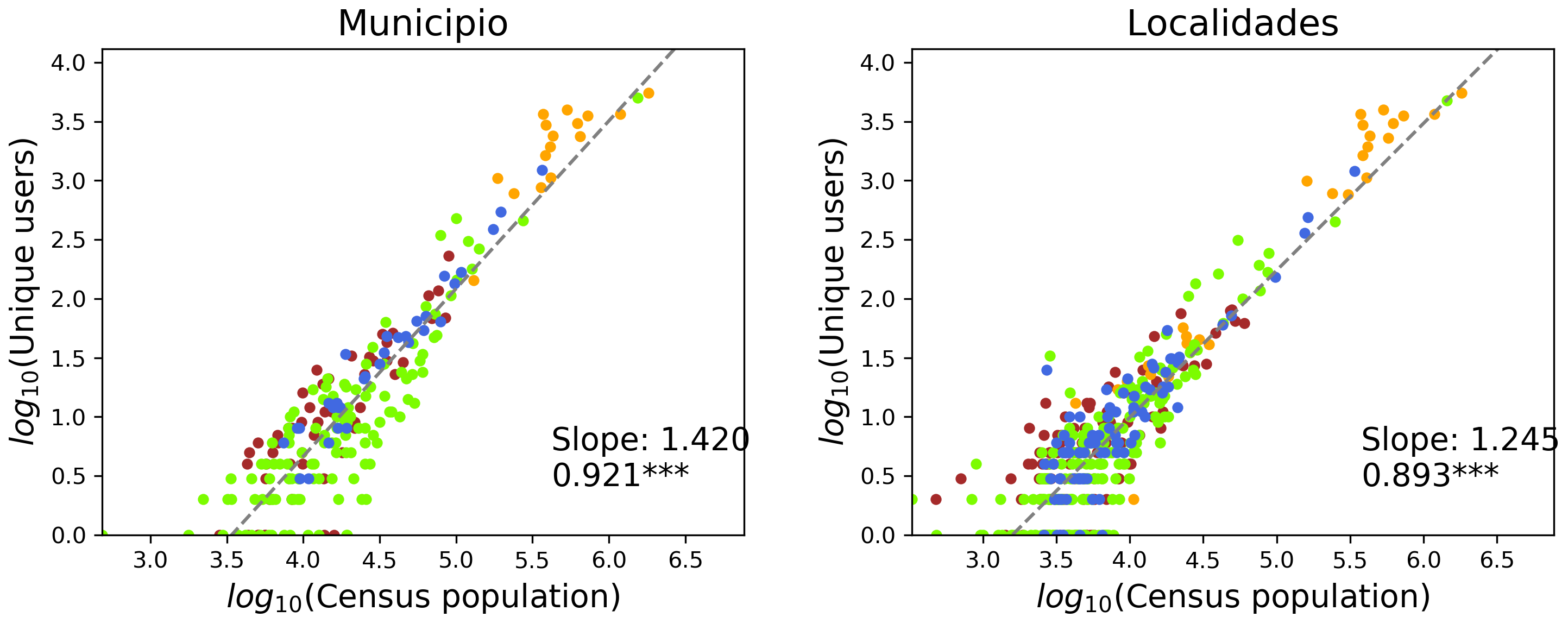}
\caption{Correlation between mobile phone data and population from census. Good correlation between mobile phone data and census information on both \textit{municipio} (US  County  equivalent) and \textit{localidades} levels. Colors represent the \textit{estadio} (states) (orange: Mexico City, green: Puebla, blue: Morelos, brown: Tlaxcala).} 
\label{samplerates}
\end{figure*}




At this stage, however, important challenges remain for the application and adoption of GPS data \cite{barra2020solid} in DRM and urban resilience efforts. Important dimensions of interest \cite{mueller2014heat,chen2018coastal}, such as household income, disaster intensity, and inequality remain challenging to overlay and link with location data. 
Technical and computational barriers to entry preclude near-real time analysis of this data.
To narrow this operational gap, in this study, we seek to elucidate a range of applications of mobility data for post-disaster analytics. We use the Puebla Earthquake as a case study (magnitude Mw 7.1, epicenter 55 km south of Puebla city), which occurred on September 19, 2017 in Mexico, causing the collapse of more than 40 buildings, displacing over 5000 people and affecting over 34,000 homes \cite{alcocer2018advisory} in and around Greater Mexico City. For each analytical finding, we present how \textit{Mobilkit}, a Python library purpose-built for DRM analytics using location data, can be leveraged for for scalable, replicable analysis in other DRM contexts.

\section{Data and Methods}

\subsection{Data}
\subsubsection{\textbf{Smartphone location data.}}
Anonymized and privacy-enhanced smartphone location data was provided by Cuebiq Inc, which collects first-party data from users who opt in through a GDPR-compliant framework (\url{https://www.cuebiq.com/}). The data spanned a 5-week period from the last week of August to end September, 2017. The Puebla earthquake occurred 2 weeks into the dataset. 
The dataset contained 774,343 anonymized unique user ids, each with a variable number of observations throughout the period. To ensure a reliable signal, the analysis was conducted on users with high observation rates: users with an average of less than three observations per day were excluded, as well as those who were observed for less than 5 days at their estimated home locations during the 14 day period prior to the earthquake. 

\subsubsection{\textbf{Socio-demographic data.}}
Population data was provided by Mexico's National Institute of Statistics and Geography (INEGI) via their microdata portal (\url{https://www.inegi.org.mx/datos/default.html#Microdatos}). Since median household income data was not available for Mexico, a composite measure of average values for asset ownership, such as vehicles, refrigerators, and computers, as well as access to services such as education and health, was created at the \textit{manzana} level (the smallest administrative unit). This multitude of features was then reduced to a single index via principle component analysis (PCA) \cite{fraiberger2020uncovering,vyas2006constructing}. We refer to this index as the ``\textit{poverty index}'' in this study. 

\subsubsection{\textbf{Seismic intensity data.}}
A seismic intensity map produced by the United States Geological Survey\footnote{\url{https://earthquake.usgs.gov/earthquakes/eventpage/us2000ar20/shakemap/}} was used to observe exposure. Seismic intensities of greater than 7.0 were observed in Puebla and parts of greater Mexico City. By overlaying these geographic extents with administrative boundaries, seismic intensities were mapped to each administrative unit.

\subsection{Methods}

\begin{figure*}[t]
  \centering
  \includegraphics[width=\textwidth]{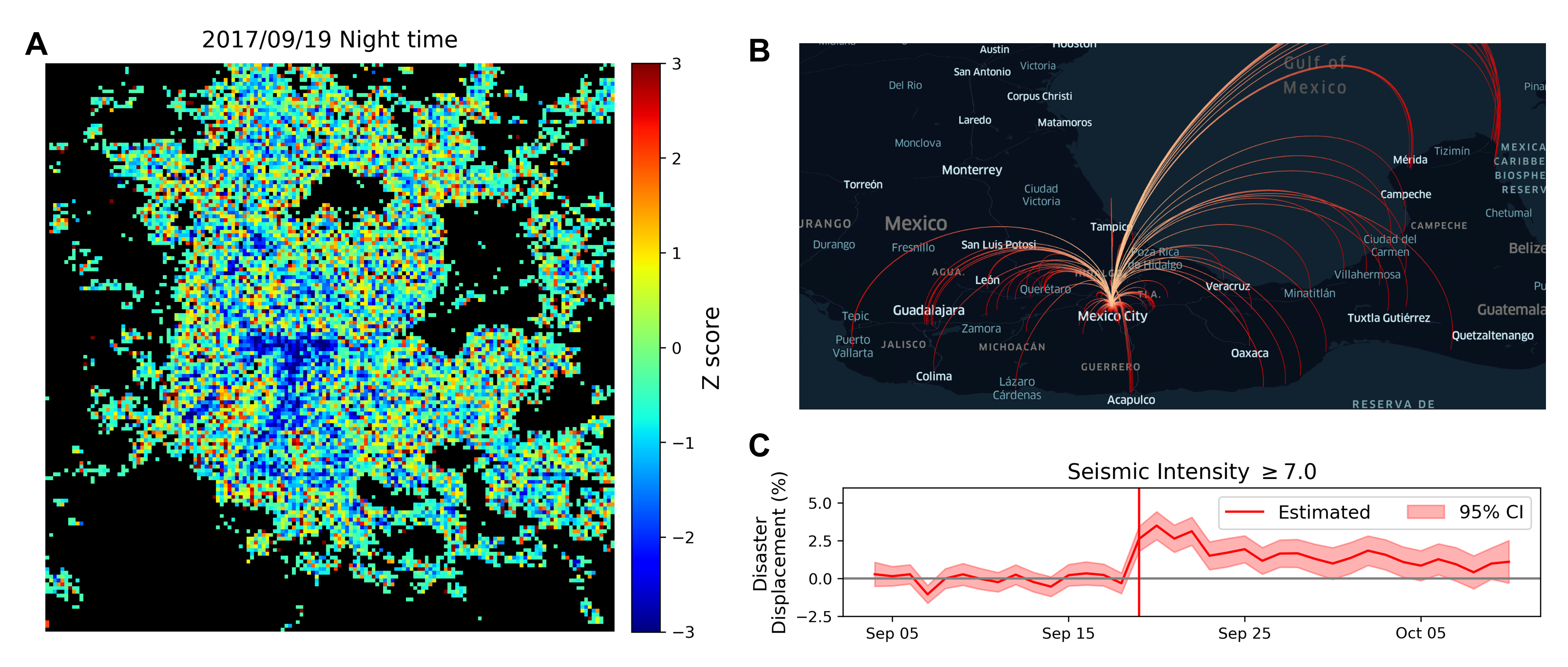}
  \caption{Displacement patterns. A) Mapping of the anomaly in population density during the night on the day of the Puebla Earthquake (September 19th, 2017) in Mexico City. Blue areas indicate decrease in active population after the earthquake. B) Displacement destinations of individuals with home locations in central Mexico City. (visualized using \url{kepler.gl}. C) Displacement recovery curve of individuals affected by seismic intensity larger than 7.0.}
  \label{displacement}
\end{figure*}

\subsubsection{\textbf{Home location estimation.}}
Home locations (assuming one primary location for each user) were detected by applying the mean-shift clustering algorithm \cite{cheng1995mean, calabrese2011estimating} to nighttime stay points observed between 8PM and 6AM. The algorithm requires 2 parameters; the density function of the kernel (Gaussian), and the kernel bandwidth (we used 200m).

\subsubsection{\textbf{Data representativeness analysis.}}
In order to quantify potential sampling bias in the data, the correlation between the number of mobile phone users and corresponding census populations at \textit{municipio} and \textit{localidades} levels was calculated. Figure \ref{samplerates} illustrates a strong positive correlation between the number of mobile phone users and census population data on both the municipio and localidades levels, (\textit{R-squared}=0.9, showing good fit; *** denotes $p<0.01$). 

\subsubsection{\textbf{Mobilkit.}}
A series of helper functions was developed to facilitate the analysis below, and packaged into \textit{Mobilkit}, a Python library purpose-built for DRM analytics using location data. Further details are provided in an upcoming paper \cite{mobilkit}. The code is open-source, released under the MIT license and can be accessed on Github (\url{https://github.com/GFDRR/mobilkit}), with complete documentation, examples and tutorials in the form of jupyter notebooks (\url{https://mobilkit.readthedocs.io/en/latest/}). Mobilkit can be easily installed on the command line, with \texttt{pip install mobilkit}.
\textit{Mobilkit} hosts a wide range of functions to conduct pre-processing and validation of a GPS location dataset before jumping into the analysis. 
For example, functions in the \texttt{stats.py} class such as \texttt{userStats()} compute statistics for each user, including the number of days spanned (time from first to last ping), number of days active (actual number of days being active), number of pings per user, and number of pings per user per active day. 
\texttt{plotUsersHist()} allows us to plot the 2-dimensional histogram of the user selection threshold parameters and the number of selected users in the dataset.

\begin{figure*}[t]
\centering
\includegraphics[width=.6\linewidth]{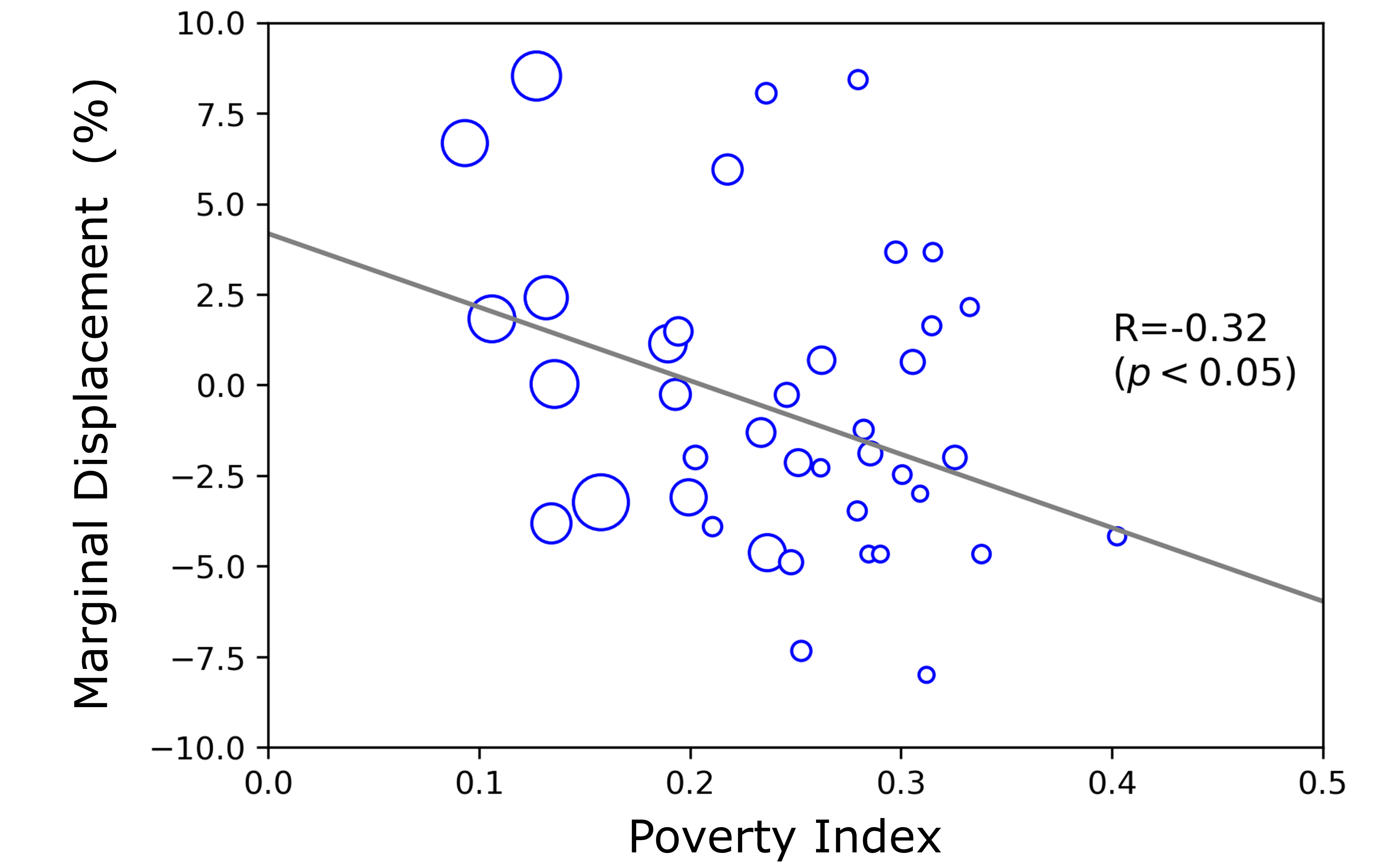}
\caption{Correlation between the poverty index and displacement rates. Negative and significant ($p<0.05$) correlation indicates that wealthier users had higher likelihood of being displaced from the affected areas. Size of the plots indicate the sample size in each municipio.} 
\label{wealth}
\end{figure*}

\section{Results}

\subsection{Quantifying displacement using anomaly detection methods} 
In order to visualize population displacement patterns after disaster events, we can map the \textit{z-scores} of the nighttime population in Mexico City on the day of the earthquake (September 19th, 2017), as shown in Figure \ref{displacement}A.
The z-score is defined as the number of standard deviations from the pre-disaster mean population, for each grid cell. We observe blue-colored clusters (\textit{z}$ <-2$) in central Mexico City, indicating significant population displacement. 
Figure \ref{displacement}B visualizes the destinations of those users displaced from central Mexico City using \url{kepler.gl}. 
While most users stay close to Mexico City, some users travel 100s of kilometers to further regions such as Cancun and Acapulco. 

In order to reproduce this analysis using \textit{Mobilkit}, the \texttt{calc\_displacement()} function can be used to compute (i) the average and minimum distances from home, (ii) the closest point to home, and (iii) the original home location of the user on each day for each user. 
Functions in the \texttt{viz} class, such as \texttt{plot\_density\_map()}, can be used to spatially plot the population density and their anomaly values in user-identified spatial granularity and scale. 

\subsection{Quantifying Displacement Rates and Recovery} 
The displacement distance of a user is defined as the minimum distance from home during the nighttime, and the user is labeled as `displaced' if the minimum displacement distance is further than the distance threshold (500 meters). 
Figure \ref{displacement}C shows that around 4\% of the users that experienced a seismic intensity greater than 7.0 were initially displaced due to the disaster, and the impact persisted for more than 2 weeks after the disaster.  
Similarly to other disasters studied in the past including Hurricanes Maria, Irma, and the Tohoku Tsunami, we observe a gradual, exponential-like decrease in population displacement \cite{yaberecovery}.

Furthermore, as shown in Figure \ref{wealth}, analysis showed a statistically significant positive correlation between wealth (poverty index) and displacement rates, indicating inequality in post-disaster displacement mobility, where poorer communities are less likely to move out of disaster affected areas than wealthier communities. 
Such analysis on displacement rates over time, and displacement rates across different socio-demographic and income groups, can be conducted using functions in the \texttt{stats.py} class of \textit{Mobilkit}. 
The documentation of \textit{Mobilkit} demonstrates how different datasets (e.g., income data and administrative boundary data) are spatially joined and processed to produce these plots.

\begin{figure*}[t]
\centering
\includegraphics[width=\linewidth]{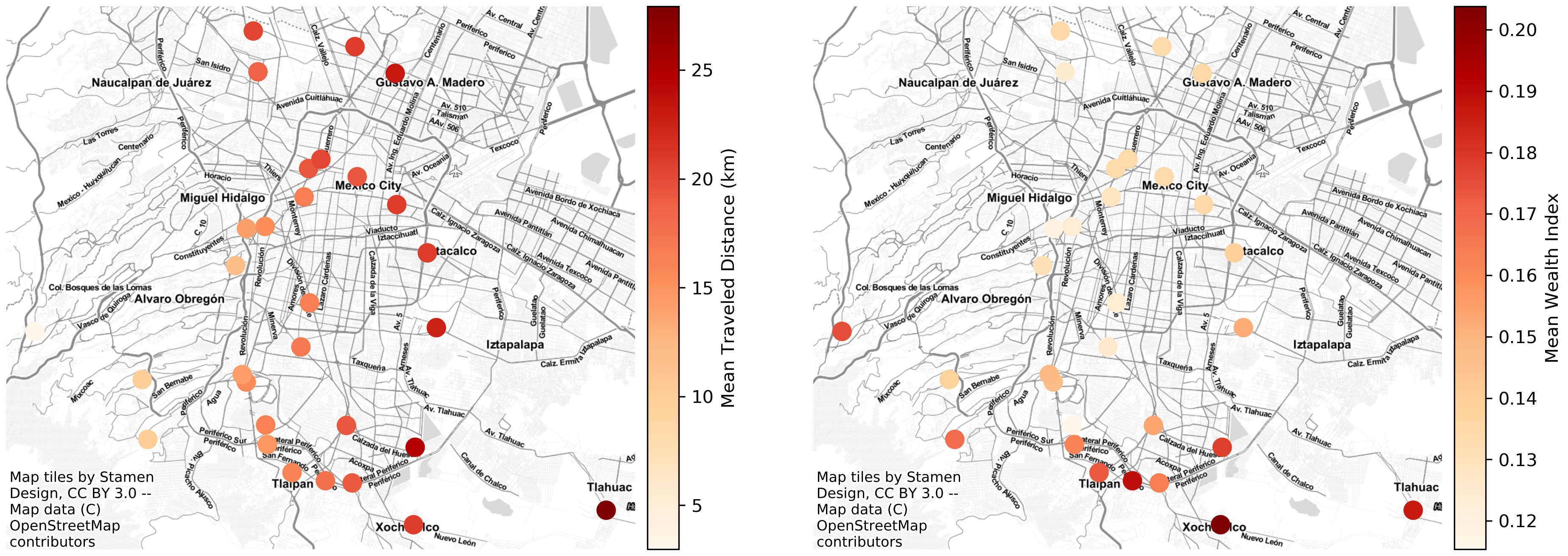}
\caption{User catchment of gathering locations. Left: mean traveled distance of users who visited each gathering location in CDMX. Right: mean poverty index value of users (estimated based on their home municipio) who visited each gathering location in CDMX.} 
\label{usercatchment}
\end{figure*}

\noindent
\subsection{Post-Disaster Gathering Locations}
To understand catchment characteristics of gathering locations after disasters, we analyze the profiles of users who visited each gathering location. 
Figure \ref{usercatchment} shows the differences in user profiles across gathering locations. 
The left panel shows that the distance traveled varies, and that gathering locations in east CDMX tend to have longer travel distances, indicating poor shelter accessibility. 
The right panel shows the mean poverty index value of users who visited each gathering location in CDMX.
There is strong spatial correlation, where gathering locations in southern CDMX accommodate residents of poorer areas. 
Such information could be useful for decision makers to develop strategies for the allocation of emergency goods and services. 

The number of visits to different points-of-interest (POIs) can be computed using raw mobility data using functions in the \texttt{spatial.py} class, such as the \texttt{compute\_poi\_visit()} function, which utilizes the spatial tree algorithm to efficiently calculate the set of users and number of users visiting a given POI for each time period identified by the user.

\section{Discussion}
In this paper we illustrate the utility of increasingly available `big data' from smartphones for post-disaster mobility analytics, such as quantifying displacement, evacuation, long-term migration, scale of impact, and more. 
This approach can be leveraged for a variety of disaster contexts beyond earthquakes that can affect human movement, such as flood impact mapping, wildfires, and more. The methods we use to detect change in population density, conduct disaster displacement analysis at varying levels of regional scale, and map movement to POIs, can be leveraged at each stage of the disaster risk management cycle.
In initial emergency phases, policymakers and local officials can leverage these methods to localize hotspots and deploy more targeted aid. Through the recovery phase, mobility data can be used to track displacement rates in time, to quantify variations in recovery and dynamically track medium-term impact. In later stages, officials can leverage long-term displacement and migration patterns in urban planning and resilience efforts, to build new shelters, reconstruct affected areas, and 'build back better'.

\begin{acks}
We extend our sincere gratitude to Cuebiq for providing the data to support this effort, and Purdue University and Mind Earth for their contribution. This work has been conducted under a grant from the Spanish Fund for Latin America and the Caribbean (SFLAC) under the Disruptive Technologies for Development (DT4D) initiative at the World Bank. The findings, interpretations, and conclusions expressed in this paper are entirely those of the authors. They do not necessarily represent the views of the International Bank for Reconstruction and Development/World Bank and its affiliated organizations, or those of the Executive Directors of the World Bank or the governments they represent. 
\end{acks}

\bibliographystyle{ACM-Reference-Format}
\bibliography{sample-base}


\end{document}